\documentclass[useAMS,usenatbib]{mn2e}

\usepackage[ansinew]{inputenc}
\usepackage{graphicx} 
\usepackage{xcolor,ulem}

\def \mnras {MNRAS}
\def \apj {ApJ}
\def \apjs {ApJS}

\def \aap {A\&A}
\def \aaps {A\&AS}
\def \nat {Nature}

\def \rmxaa {RMxAA}
\def \pasa {PASA}
\def \physscr {Physica Scripta}

\def \angstron {$\rm \AA$}
\def \teff {T$_{\rm eff}$}

\def \logg {$\log{g}$}

\title[ALiCCE]{ALiCCE: Atomic Lines Calibration using the Cross-Entropy Algorithm}
\author[Martins et al.]{Lucimara P. Martins$^{1}$\thanks{E-mail:
lucimara.martins@cruzeirodosul.edu.br}, Paula Coelho$^{1,2}$, Anderson Caproni$^{1}$, Roberto Vitoriano$^{1}$\\
$^{1}$NAT - Universidade Cruzeiro do Sul, Rua Galv\~ao Bueno, 868, S\~ao Paulo, SP, Brazil\\
$^{2}$IAG - Universidade de S\~ao Paulo, R. do Mat\~ao, 1226, 05508-090,S\~ao Paulo, SP, Brazil}
\begin{document}

\date{Accepted ? December ? Received ? December ?; in original form ? October ?}

\pagerange{\pageref{firstpage}--\pageref{lastpage}} \pubyear{2009}

\maketitle

\label{firstpage}

\begin{abstract}

Atomic line opacities play a crucial role in stellar astrophysics.
They strongly modify the radiative transfer in stars, therefore impacting
their physical structure. Ultimately, most of our knowledge of stellar population
systems (stars, clusters, galaxies, etc.) relies on the accuracy with which we understand
and reproduce the stellar spectra. 
With such a wide impact on Astronomy, it would be ideal to have access to a complete, 
accurate and precise list of atomic transitions. This, unfortunately, is not the case.
Few atomic transitions had their parameters actually
measured in the laboratory, and for most of the lines the parameters were calculated with low precision 
atomic energy levels. Only a small fraction of the lines were calibrated empirically. 
For the purpose of computing a stellar spectral grid with a complete coverage of spectral types and luminosity classes, 
this situation is rather limiting.
We have implemented an innovative method
to perform a robust calibration of atomic line lists used by spectral synthesis
codes called 
ALiCCE: Atomic Lines Calibration using the Cross-Entropy algorithm. Here we describe the implementation and validation 
of the method, using synthetic spectra which simulates the signal-to-noise, spectral resolution and rotational velocities 
typical of high quality observed spectra. We conclude that the method
is efficient for calibrating atomic line lists.

\end{abstract}


\section{Introduction}

Atomic line opacities play a crucial role in stellar astrophysics. First,
they strongly modify the radiative transfer in stars, and therefore impact
their physical structure. Secondly, the detailed comparison
between model and observed spectra is a powerful
diagnostic tool that may be used to study stars and stellar populations.
Everything we know about the chemical abundances on stars, planets, galaxies, 
interstellar, intergalactic and intracluster medium is, ultimately, dependent 
on the quantum mechanic parameters which characterize atomic (and molecular) 
electronic (vibrational and rotational) transitions. These parameters will, 
given the thermodynamic conditions in which the transition happens, define the 
profile of the spectral line. For the case of atomic transitions, these parameters are:
the central wavelength of the line, the energy levels of the transition, the oscillator
strength (which dominates the line depth) and the broadening parameters 
(which dominate the wings of the lines).

With such a wide impact on Astronomy, one would expect by now to have access to a complete, 
accurate and precise list of atomic transitions. This, unfortunately, is not the case. 
Although nowadays there exist fairly comprehensive databases (e.g., 
the NIST Atomic Spectra Database~\footnote{Available at \\
http://www.nist.gov/pml/data/asd.cfm.}
and the Vienna Atomic Line Database~\footnote{Available at 
http://www.astro.uu.se/~vald/php/vald.php})
of the atomic transition probabilities necessary for stellar abundance studies, 
they are far from being complete.

In general, one half the discernible lines in observed stellar spectra are missing from the line lists with
good wavelengths \citep{kurucz11}. 
To have an accurate opacity list for atoms and ions, we need all levels, including hyperfine and isotopic 
splittings. 
Lifetimes and damping constants depend on sums over the levels. Inside stars there are 
thermal and density cutoffs that limit the number of levels, but in circumstellar, interstellar, 
and intergalactic space, photoionization and recombination can populate high levels, 
even for high ions. We need all stages of ionization for elements at least up through Zn.
In the Sun there are unidentified asymmetric triangular features that are unresolved multiplets of 
light elements \citep{kurucz11}.
All the magnetic dipole, electric quadrupole, and maybe higher-pole, forbidden lines are required as well. 
However, except for the simplest species, it is impossible to generate 
accurate energy levels or wavelengths theoretically. In principle,
they must be measured in the laboratory. 

Because the parameters of relatively few lines were actually measured in laboratory, 
to compute a theoretical stellar spectra
with good spectrophotometry is necessary to include the
so-called ``predicted lines'': lines where either one or both energy
levels of the transition were predicted from quantum mechanics calculations \citep{kurucz92}. 
Usually only the lower energy
levels of atoms have been determined in the laboratory, particularly for complex spectra 
such as those from iron. If only those
transitions were taken into account, the atmospheric line blanketing computed 
from such data would be severely incomplete.
The predicted lines are essential for computing accurately the
structure of model atmospheres and for spectrophotometric predictions 
\citep[e.g.][]{short+96}. But as the quantum mechanics predictions are accurate to 
only a few per cent, wavelengths
for these lines may be largely uncertain. Also the line oscillator
strengths are sufficiently accurate merely in a statistical sense.
The predicted lines are, therefore, unsuitable for high resolution
analyses \citep{bell+94, castelli+04,munari+05}. 

Efforts are underway to reduce transition probability uncertainties of selected lines 
\citep[e.g.][]{taklif90,klose+02,fuhr+06, safronova+10,
baclawski11,pickering+11,wiese+11,civis+12,ruffoni+13} 
and accurately compute broadening parameters 
\citep[e.g.][]{anstee+95,barklem+97,barklem+98,lesage+99,barklem+00,konjevic+02,
derouich+03,dimitrijevic+03}.
Also, for the purpose of high spectral resolution chemical analysis of stellar photospheres, 
several limitations of the atomic line lists are being successfully tackled by different authors and methods 
\citep[e.g.][]{stalin+97,blackwell+08,borrero+03,
jorissen04,sobeck+07,melendez+09,wahlgren10,denHartog+11,shchukina+13,wood+13}.

It has also been shown in the literature \citep[e.g.][]{barbuy+03,martins+07} that even empirical calibrations
of some specific lines can produce significant improvement on the synthetic spectra
generated. The empirical calibration is done by changing the values of the parameters on the line list, 
generating models and comparing them with
observations of very well known stars (like the Sun or Arcturus, for example). 
This process is repeated until the results are adequate \citep{barbuy+03}.

Nevertheless, these approaches tend to improve the quality of a selective group lines, 
mostly those that were considered more suitable to chemical abundance measurements, where relatively weak lines in the linear part
of the curve of growth are favored. On the other hand, the strong lines close to saturation and blended features are the ones which dominate spectral 
indices in integrated spectra of stellar populations.

While these efforts are improving the parameters for thousands of lines, 
tens of millions of lines are estimated to be needed to compute, say, a complete stellar 
grid with a good range of atmospheric parameters.
Therefore, an innumerous amount of lines remain poorly characterized. 
For the ultimate goal of computing a large grid of theoretical stellar 
spectra for further use in automatic classification of stellar spectroscopic 
surveys and stellar population modeling, this is rather limiting.

Aiming to overcome these limitations and to create a more robust and automatic method
for the calibration of atomic parameters,
we introduce a powerful statistical technique recently developed to deal with multi-extremal
problems involving optimization: the cross-entropy algorithm (hereafter, CE).

In this work we validate our CE method to calibrate atomic line lists
using synthetic spectra which simulates very well known stars. 
The code is hereafter called
ALiCCE: Atomic Lines Calibration using the Cross-Entropy algorithm.
We show the great capability
of the method to determine the atomic parameters of the absorption lines by comparing
the models with the simulated spectra. This paper is structured as follows: 
in \S 2 we explain the Cross-Entropy method, in \S 3 we explain the methodology
used in the validation process, in \S 4 we show our results and
in \S 5 we present our discussion and conclusion.

\section{The Cross-Entropy Method and ALiCCE}

To compute a synthetic stellar spectrum, the spectral synthesis code reads a model atmosphere, and
a list of molecular and atomic transitions. From the many model atmospheres available in the literature 
we choose the library from \citep{castelli+04}
which is based on ATLAS9\footnote{Available at \\http://wwwuser.oats.inaf.it/castelli/sources/atlas9codes.html}
\citep{kurucz70,sbordone+04}. This is an extensive grid of models and according to 
\citet{martins+07} it is one that in average, reproduces well the colors of
observed stars. 

To generate the synthetic spectra we chose the code SYNTHE \citep{kurucz+81}
in its public Linux port by \citet{sbordone+04}. 
SYNTHE reads the atomic and molecular transition lines in
separated files. The atomic lines are listed in files which are publicly distributed with 
SYNTHE. 

The synthesis codes use the atomic line lists to solve the continuum radiative transfer equation 
and to determine the absorption lines formation. The parameters needed for the computation
of each line profile are: the central wavelength, the transition levels energy, the oscillator
strength $\log(gf)$ (which is related to the line intensity) and the broadening parameters 
(natural - $\log ~\Gamma_{rad}$, Stark - $\log~\Gamma_{StarK}$ and Van der Waals - $\log \Gamma_{vdW}$ broadening). 
Broadening parameters are responsible for the line profile. The natural broadening
comes directly from the Heisenberg's uncertainty principle of quantum mechanics:
because an electron that finds itself in an energy level has a finite lifetime
before transitioning to a lower energy level, the energy levels have a 
certain width $\Delta E$, resulting in a width of the atomic line. The
radiative damping constant $\Gamma_{rad}$ is then equal to the sum 
of the reciprocal of the mean lifetimes of the 
two atomic energy levels under consideration. Another mode of line
broadening is the pressure (or collisional) broadening, which is
due to the perturbation of the potential of the atom by neighboring particles.
For example, the Stark effect is caused by the splitting of degenerate atomic energy levels
due to the presence of an external electric field, or a frequency shift
for these levels. Another example of pressure broadening is the Van der Waals process,
which is related to the perturbation of an
atom's potential by neutral atoms electric dipole. Their definition implies 
that the Stark process will be more important for hotter stars while
the Van der Waals will be more important for cooler stars.

In the present work we choose to calibrate the 
oscillator strength and the two pressure broadenings - Stark and
Van der Waals.

Another detail we have to consider to calibrate the lines is that 
although the same line list is used to generate the spectra of
all stars, not every line will be 
present in all stellar spectra. 
The intensity and the profile of each absorption feature is determined
by the physical conditions of the stellar atmosphere where the lines
are formed. 
Lines originated from the lowest ionization degrees will be more
intense in cooler stars. On the other hand, lines originated
from higher ionization degrees will be stronger on the spectrum
of hotter stars.
Many other 
physical parameters can also change the intensity and the profile of
the lines, like chemical abundances, gravity, etc. 
Besides, many lines are blended in
the spectrum of a star, which makes it very hard to disentangle the atomic parameters
of each line. 
Ideally, in order to be able to calibrate all the lines in a given line list
one would have to attempt to calibrate the lists
to a variety of stellar spectral types simultaneously.
In different spectral types, the
relative intensity of each line in a blend might be different. Calibrating
the line list to all of them simultaneously would better constraint the best solution.

\subsection{The Cross-Entropy Algorithm}

For the task of calibrating the atomic line list we adapted the CE algorithm. 
The CE method is a general Monte Carlo approach to combinatorial and
continuous multi-extremal optimization and importance sampling, which is
a general technique for estimating properties of a particular distribution,
using samples generated randomly from a different statistical distribution rather
than the distribution of interest. 

The CE analysis was originally used in the optimization of complex 
computer simulation models involving rare
events simulations \citep{rubinstein97}, where very small probabilities
have to be accurately estimated, having been modified
by \citet{rubinstein99} to deal with continuous multi-extremal
and discrete combinatorial optimization problems. 
Its theoretical asymptotic convergence has been demonstrated
by \citet{margolin04}, while \citet{kroese+06} studied the efficiency of the CE method in solving
continuous multi-extremal optimization problems. Some examples of 
robustness of the CE method in several situations
are listed in \citet{deboer+05}. The CE procedure uses concepts of importance sampling,
 which is a variance reduction technique, but removing the need for a
priory knowledge of the reference parameters of the parent 
distribution. The CE procedure provides a simple adaptive way of
estimating the optimal reference parameters. 

The CE method was already used
successfully in astrophysical problems like the study of sources with
jet precession \citep{caproni+09, caproni+13}, fitting the color-magnitude
diagram of open clusters \citep{monteiro+10,oliveira+13} and modeling very
long baseline interferometric images \citep{caproni+11}.

The basic procedures involved in the CE optimization can be summarized as follows
\citep[e.g.][]{kroese+06}:

   (i) Random generation of the initial parameter sample,
obeying predefined criteria;

   (ii) Selection of the best samples based on some mathematical criterion;

   (iii) Random generation of updated parameter samples
from the previous best candidates to be evaluated in the
next iteration;

   (iv) Optimization process that repeats steps (ii) and (iii) until
a pre-specified stopping criterion is fulfilled.

Let us suppose that we wish to study a set of $N_\rmn{d}$ observational 
data in terms of an analytical model characterized by $N_\rmn{p}$ 
parameters $p_1, p_2, ..., p_{N_\rmn{p}}$.

The main goal of the CE continuous multi-extremal optimization 
method is to find the set of parameters ${\bf x}^*=(p^*_1,p^*_2,...,p^*_{Np})$ for 
which the model provides the best description of the data \citep{rubinstein99,
kroese+06}. 
In our case $N_\rmn{d}$ is the number of stellar spectra we are trying to
reproduce, and ${N_\rmn{p}}$ is three times the number of lines in the
interval calibrated (because we are calibrating three parameters for 
each line).
The technique is performed by generating 
randomly $N$ independent sets of model parameters 
${\bf X}=({\bf x}_1,{\bf x}_2,...,{\bf x}_N)^T$, 
where ${\bf x}_i=(p_{1i},p_{2i},...,p_{N{\rmn{p}i}})$, and minimizing an 
objective function $S({\bf x})$ used to transmit the quality of the fit 
during the run process. If the convergence to the exact solution is 
achieved then $S({\bf x}^*)\rightarrow 0$.

In order to find the optimal solution from CE optimization, 
we start by defining the parameter range in which the algorithm 
will search for the best candidates: $p^\rmn{min}_j\leq p_j(k) \leq p^\rmn{max}_j$, 
where $k$ represents the iteration number. 
Introducing $\bar{p}_j(0)=(p^\rmn{min}_j+p^\rmn{max}_j)/2$ and $\sigma_j(0)=(p^\rmn{max}_j-p^\rmn{min}_j)/2$, 
we can compute ${\bf X}(0)$ from:

\begin{equation}
  X_{ij}(0)=\bar{p}_j(0)+\sigma_j(0) G_{ij},
\end{equation}
where $G_{ij}$ is an $N\times N_\rmn{p}$ matrix with 
random numbers generated from a zero-mean normal distribution 
with standard deviation of unity.

The next step is to calculate $S_i(0)$ for each set of ${\bf x}_i(0)$, 
ordering them according to increasing values of $S_i$. 
Then the  first $N_\rmn{elite}$ set of parameters 
is selected, i.e. the $N_\rmn{elite}$-samples with lowest $S$-values, 
which will be labeled as the elite sample array ${\bf X}^\rmn{elite}(0)$.

We then determine the mean and standard deviation of 
the elite sample, $\bar{p}^\rmn{elite}_j(0)$ and ${\bf \sigma}^\rmn{elite}_j(0)$ 
respectively, as:

\begin{equation}
  \bar{p}^\rmn{elite}_j(0)=\frac{1}{N_\rmn{elite}}\sum\limits_{i=1}^{N_\rmn{elite}}X^\rmn{elite}_{ij}(0),
\end{equation}

\begin{equation}
  {\bf \sigma}^\rmn{elite}_j(0)=\sqrt{\frac{1}{\left(N_\rmn{elite}-1\right)}\sum\limits_{i=1}^{N_\rmn{elite}}\left[X^\rmn{elite}_{ij}(0)-\bar{p}^\rmn{elite}_j(0)\right]^2}.
\end{equation}

The array ${\bf X}$ at the next iteration is determined as:

\begin{equation}
  X_{ij}(1)=\bar{p}^\rmn{elite}_j(0)+{\bf \sigma}^\rmn{elite}_j(0) G_{ij},
\end{equation}

This process is repeated from equation (2), with $G_{ij}$ 
regenerated at each iteration. The optimization stops when either 
the mean value of ${\bf \sigma}^\rmn{elite}_i(k)$ is smaller than a 
predefined value or the maximum number of iterations $k_\rmn{max}$ is reached.
In our case the stopping criteria is always $k_\rmn{max}$, as will be explained in 
section 4. 

In order to prevent convergence to a sub-optimal solution due to the intrinsic 
rapid convergence of the CE method, \citet{kroese+06} suggested the implementation 
of a fixed smoothing scheme for 
${\bf \sigma}^\rmn{elite,s}_j(k)$:


\begin{equation}
  {\bf \sigma}^\rmn{elite,s}_j(k)=\alpha_\rmn{d}(k){\bf \sigma}^\rmn{elite}_j(k)+\left[1-\alpha_\rmn{d}(k)\right]{\bf \sigma}^\rmn{elite}_j(k-1),
\end{equation}
where $\alpha_\rmn{d}(k)$ is a 
dynamic smoothing parameter at $k$th iteration:

\begin{equation}
  \alpha_\rmn{d}(k)=\alpha-\alpha\left(1-k^{-1}\right)^q,
\end{equation}
with $0<\alpha< 1$ and $q$ is an integer typically between 5 and 10 
\citep{kroese+06}.

As mentioned before, such parametrization prevents the algorithm 
from finding a non-global minimum solution since it guarantees 
polynomial speed of convergence instead of exponential \citep{kroese+06}. 

\subsection{The performance function}

In order to select the best model that represents our observation
we need to define a performance function, based on the desired characteristics
of the solution. Usually, for continuous problems this is done by defining a 
likelihood function and then maximizing it, or alternatively requiring
that the sum of the squared residuals be minimal.

In this work the performance function has to be defined
as a way to compare model with observed spectra pixel by pixel.
For this comparison  we choose as a performance function
the combination of the sum of the squared  residuals and their respective
variance, as suggested in \citet{caproni+11}.

Let the quadratic residual $R_m(k)$ at a given wavelength pixel $m$ and
iteration $k$ be defined as the squared difference between the observed
spectrum we want to reproduce, $I_m$, and the generated model spectrum $M_m(k)$ at a 
$k$-iteration, i.e. $R_m(k) = [I_m - M_m(k)]^2$. The mean square
residual value of the model fitting $\bar{R}(k)$ can be 
calculated from:

\begin{equation}
\bar{R}(k)=\frac {1} {N_\mathrm{pixel}}\left[\sum\limits_{m=1}^{N_\mathrm{pixel}} R_m(k)\right].
\end{equation}

As mentioned in the previous section, we need the spectra of
stars spanning a range of effective temperatures (\teff), surface gravities(\logg)
and metallicity
in order to guarantee that lines to be calibrate will 
be present in at least one spectrum. Considering $N_\mathrm{s}$ as the number of stars used in 
the calibration, our tentative model spectra obtained from the 
$N_\mathrm{s}$ parameters ${\bf x}_i(k)$ at iteration $k$ are ranked
through the performance function:
 
\begin{equation}
  S_\mathrm{prod}({\bf x}_i,k)=\prod_{j=1}^{N_S} \bar{R}(k)\times\frac {1} {N_\mathrm{pixel}}
  \left[\sum\limits_{m=1}^{N_\mathrm{pixel}} \left(R_m(k)-\bar{R}(k)\right)^2\right],
\label{perffunc}
\end{equation}

which corresponds to the product of individual merit functions
associated with the $N_\mathrm{s}$ stars used in the calibration.
The product of these functions is important to ensure that high luminosity stars do not become
more important than low luminosity stars. 
Note also that, if $I_m$ and $M_m$ are expressed in terms of ergs cm$^2 \angstron^{-1}$, 
$S_\mathrm{prod}$ has units of (ergs cm$^2 \angstron^{-1})^{6N_S}$.

It is important to emphasize that we have tested other functional forms 
for the CE performance function (as a root mean square, for example)
 but Equation (8) was more efficient to recover the best solution
for our problem. As mentioned by \citet{caproni+11}, the difference in performance 
is essentially due to the fact that Equation (8) transmits directly 
any change in the mean and variance of the residuals in all iteration 
steps of the optimization process.

\subsection{Determination of the parameters and their uncertainties}

As commented previously, CE optimization generates $N$ random tentative 
solutions at each iteration $k$, selecting the best $N_\mathrm{elite}$ 
set of model parameters in terms of the values of $S_\mathrm{prod}$.

If the stopping criteria is related to a maximum number of iterations,
the last iteration might not be exactly the one with the best $S_\mathrm{prod}$. 
The best values of the parameters $p_i^*$, 
as well as their respective uncertainties $\sigma_{p_i^*}$, were then determined as follows:

\begin{equation}
p_i^*=\sum\limits_{k=1}^{k_\mathrm{conv}}w_kp_{ik}\left(\sum\limits_{k=1}^{k_\mathrm{conv}}w_k\right)^{-1}
\end{equation}
and

\begin{equation}
\sigma_{p^*_{i}}^2=\sum\limits_{k=1}^{k_\mathrm{conv}}w_k\left(p_{ik}-p^*_{i}\right)^2\left(\sum\limits_{k=1}^{k_\mathrm{conv}}w_k\right)^{-1},
\end{equation}

where $k_\mathrm{conv}$ is the iteration where the parameter converged, 
and $p_{ik}$ represents the set of model parameters that produce the minimum value 
of $S_{\mathrm{prod}_k}$ among all tentative solutions at iteration $k$ and 
$w_k=S^{-2}_{\mathrm{prod}_k}$. The power index -2 in the definition 
of $w_k$ was adopted in order to make the tentative solutions with the lowest 
values of $S_{\mathrm{prod}_k}$ more important in the calculation of $p_i^*$.
The value of $k_\mathrm{conv}$ corresponds to the iteration where the value
of $p_{ik}$ did not change after a certain number of iterations. We found that
a safe value for this number is 10\% of the total number of iterations.

\subsection{ALiCCE}

ALiCCE is a code written in C which implements the Cross-Entropy
algorithm to calibrate atomic line lists used in
stellar spectral syntehsis.
This version
of ALiCCE is adapted to make an external call to SYNTHE code for performing
the spectral syntehsis. 

For each iteration, ALiCCE generates $N$ different atomic line lists,
with each atomic parameter to be calibrated ($p_{N{\rmn{p}i}}$) varying inside a given interval.
It then calls SYNTHE for each of the stars used in the calibration (three in 
this validation case), for each one of the $N$ lists. The output spectrum generated ($M_m(k)$)
for each of the atomic line lists is then compared with the simulated observed
spectrum ($I_m$), and the performance function ($S_\mathrm{prod}$) is calculated for each of the $N$ lists.
ALiCCE then ranks the $N$ line lists by $S_\mathrm{prod}$ 
(from the lowest to the highest values)
and recalculate the interval for each atomic parameter based on 
the mean and standard deviation of the $N_{elite}$ first tentative solutions in the
$S_\mathrm{prod}$ rank ($N_{elite}$ = 0.05 $N$ - see section 3.2).
The process starts again until the stopping criteria
is fulfilled.

\section{Validating ALiCCE}

\subsection{Synthetic Spectra}

The first step to calibrate the atomic line list is
to choose the wavelength range for the calibration. For this initial validation
step we choose the range from 851.0 to 853.0 nm. There are two motivations for this choice: 
(1) It is a region relatively free of strong molecular transitions, which makes the calibration of 
atomic lines more robust and (2) It is inside the wavelength region covered by the 
Gaia Mission\footnote {http://gaia.esa.int}, which will generate spectra
of millions of stars from 8470 to 8740 \AA. Our first goal 
for the calibrated list is 
to compute a stellar library for the the Gaia Mission wavelength range. 
For the selected range of only 20\AA~there are almost 290 atomic lines. 
For illustrative purposes, we show in Table~\ref{tablelines} the Kurucz line list for the first 5 lines
of our range, extracted from the gf1200.100 file from the SYNTHE package. 

It is important to realize
the amount of calculation needed to calibrate these lists. For example, in the
Kurucz line list used here, there is an average of eighteen atomic lines per angstrom, each of these
with at least three adjustable parameters. It is unfeasible to calibrate all these lines manually for
hundreds or thousands of angstroms as a stellar library would require. 

To validate the calibration technique we used synthetic spectra instead of observed ones. These 
spectra were generated with the original line list of Kurucz, therefore we know the values of the 
atomic parameters ALiCCE has to recover. To make the test more realistic we also introduced 
some random noise in these spectra. The S/N we used here is 400, the lower limit we expect 
for the observations that will be used for the calibrations (Coelho et al. in preparation 
\footnote{
These are high-resolution (R $\sim$ 40000) observations of six 
nearby stars (spectral types  K2Ib,  K2/K3III, a solar twin, F0V, A3m, and B6III) 
obtained with UVES at VLT/ESO (ID 087.B-0308(A+B)).}).
The goal is to verify if ALiCCE is able to recover the atomic parameters without any
prior knowledge about their values, only through the comparison of the spectra.

The stars that will be used in the calibration play a crucial role.
As we mentioned before, the ideal case would be to use many stars with different
spectral types. However, in practice,
this is unfeasible. The observed stellar spectra that will be used to calibrate
the atomic line list need to correspond to stars with atmospheric parameters very well
determined as model independently as possible. The spectra also need to have
extremely high resolution and a very high signal-to-noise. In practice,
there are very few available spectra in the literature that satisfy these criteria. For this validation project, we generated the synthetic spectra with the atomic 
parameters of real stars that will be used in the actual calibration: three well known stars, that have not only very well determined atmospheric parameters but also have very different spectral types: The Sun (G2V), Arcturus (K1.5III) and Vega (A0V).
Their atmospheric parameters are listed in Table~\ref{atmospar}.

\begin{figure} 
   \centering
   \includegraphics[width=8.6truecm]{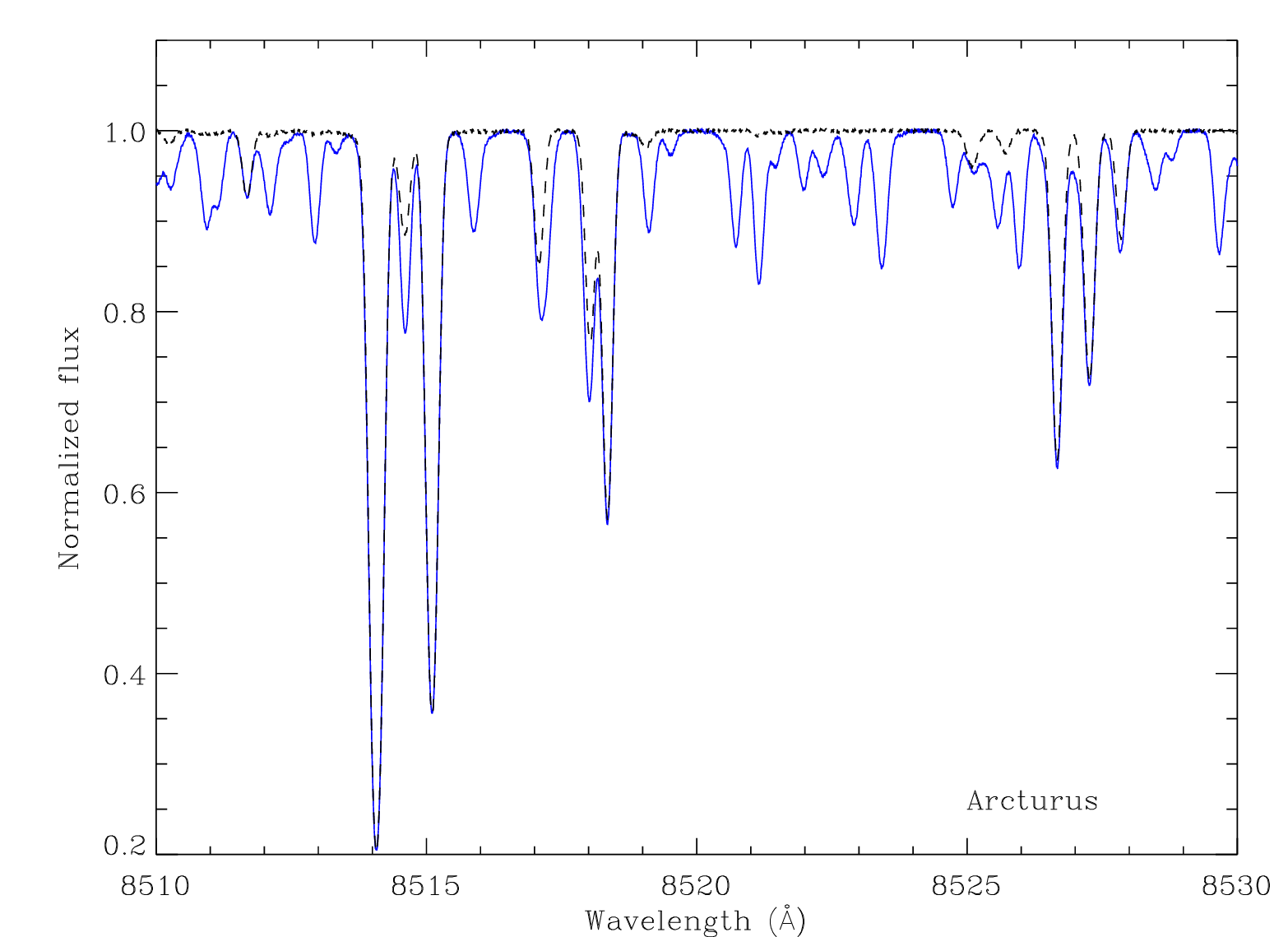} 
   \includegraphics[width=8.6truecm]{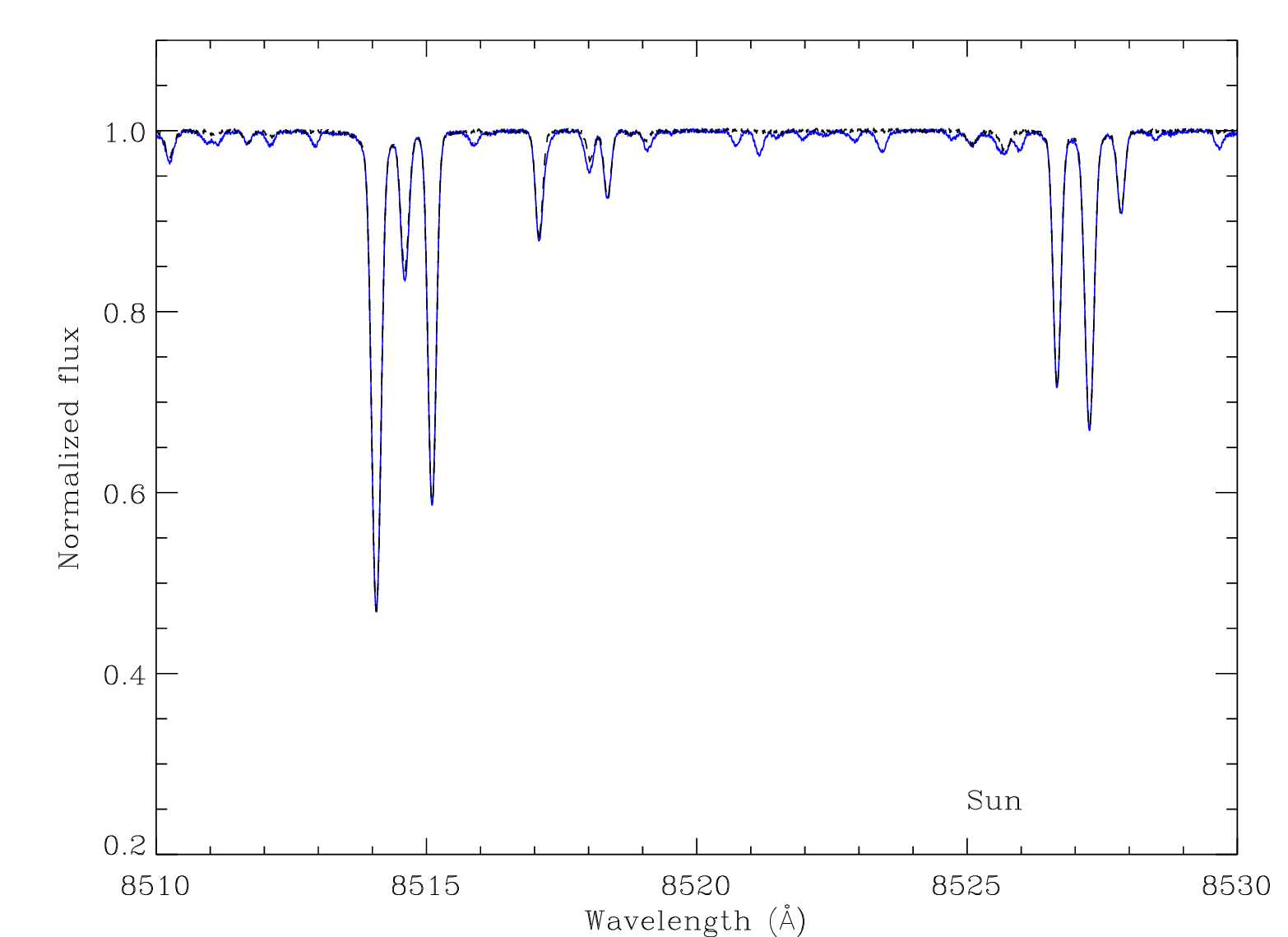} 
   \includegraphics[width=8.6truecm]{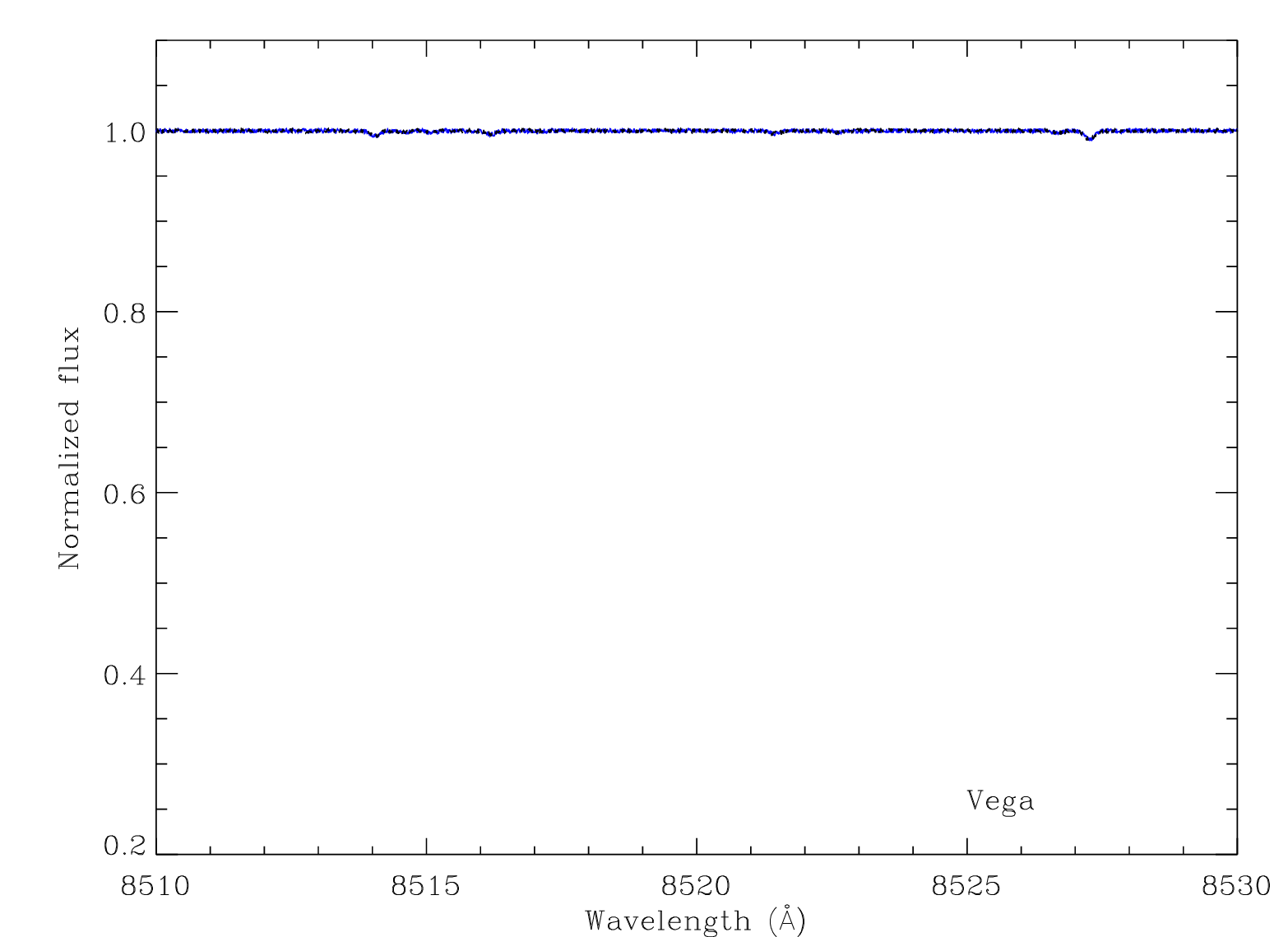} 
  \caption{Synthetic spectra used on the atomic line list calibration validation (continuous line). Noise 
was artificially added to the spectra. The S/N value adopted was 400. The figure 
also shows the spectra generated only with atomic lines (dotted line). }
\label{starsobs}
\end{figure}

The spectra used for our validation process are shown in Figure~\ref{starsobs} (continuous line). 
It is important to mention that not all the lines visible in these spectra are atomic lines. 
In this figure we also present the spectra generated only with
atomic lines (dotted line). 
Although strong molecular absorptions are not present, some of the lines in
this interval have molecular 
origin and will not be calibrated by this version of ALiCCE.

\begin{table*}
\caption{Extraction of the first 5 lines of the calibration wavelength range of the gf1200.100 file 
used in the calibration validation process.}
\begin {tabular} {@{}|c|ccccccccccc}
\hline
$\lambda$ &  Log    & Z & 1st               & J for the   & Name of the  & 2nd             & J for the  & Name of the & Log  & Log  & Log    \\
 (nm)      &   gf   &   &  Energy           & 1st Energy  & 1st Energy   & Energy          & 2nd Energy & 2nd Energy  & Rad  & Stark& VDW    \\
           &        &   &  Level  cm$^{-1}$ & Level       & Level        & Level cm$^{-1}$ & Level      & Level        &      &      &          \\
\hline

851.0171 &-1.800 & 15.00 &  67971.072 & 0.5 & 4p  2P    & 79718.490 & 1.5 & 5d  4D     & 0.00 & 0.00 & 0.00 \\     
851.0245 &-1.990 & 14.00 &  49850.830 & 2.0 & p3d  3F   & 61598.145 & 2.0 & p5f'[5]    & 0.00 & 0.00 & 0.00  \\   
851.0292 &-1.790 & 27.00 &  54946.900 & 2.5 & 5F)s4d e6D& 43199.650 & 2.5 & 4F)4sp x4G & 8.81 &-5.33 & -7.67 \\    
851.0314 &-2.398 & 25.00 &  56561.950 & 2.5 & e4D       & 44814.730 & 1.5 & (5D)4p z4F & 8.27 &-5.55 & -7.59 \\    
851.0481 &-3.740 &  6.00 &  75255.270 & 2.0 & s2p3 3P   & 87002.260 & 3.0 & p6p  3D    & 0.00 & 0.00 & 0.00 \\     

\hline
\end{tabular}
\label{tablelines}
\end{table*}

\begin{table}
\caption{Atmospheric parameters of the stars used in the validation test.}
\begin {tabular} {@{}|c|ccccc}
\hline
Star    & T$_{eff}$ (K) & log g & [Fe/H]   & vsini (km/s)& Ref   \\
\hline
Arcturus& 4275 & 1.55 & 0.34673 & 1.85   & a,b\\
Sun     & 5777 & 4.44 & 1.00000 & 2.4 & c\\
Vega    & 9550 & 3.95 & 0.31620 & 22  & d,e\\
\hline
\end{tabular}
a- \citet{gray81}; b- \citet{melendez+03};c - \citet{smith78};
d - \citet{Castelli+94}; e - \citet{peterson+06}

\label{atmospar}
\end{table}

\subsection{Methodology and Definition of CE parameters}

The performance function was defined in a way that it compares models
(spectra generated by ALiCCE) with simulated observations (input synthetic spectra with noise),
wavelength by wavelength pixel. If too many pixels are considered at once, 
the importance of each pixel to the total performance function becomes increasingly 
small. If we are aiming to calibrate not only the strong lines, but also the weak ones, 
we cannot perform the calibration in large wavelength ranges at a time. On the other hand,
the calibration cannot be performed in wavelength ranges excessively small because some lines 
can have widths of several angstroms (sometimes even tenths of angstroms).

Besides, the number of tried solutions ($N$) as well as the number of maximum iterations ($k_{max}$),
and consequently the computation time, are directly related with the number of free parameters
to be adjusted. This means that if we choose a large wavelength range to perform the calibration
at once the problem might become unfeasible. 

The calibration can be performed in small steps of wavelengths at a time.
The only compromise that must be respected is to choose a wavelength range which is large enough 
to encompass the entire lines (center and wings) that will be calibrated.
For this calibration process we performed the calibration in steps of 1.5~\AA~width, with an 
overlap of 0.5~\AA~width from one to another.

It is also important to mention that although we are calibrating only 1.5~\AA~ at a time, 
all the spectra are always generated for a much larger wavelength range 
(at least 5~\AA~to each side) to ensure no border effect will be important.

Another constrain that has to be defined is the interval in which each
atomic parameter can vary in the first iteration. Errors in the atomic parameters can be as high as 
400\% \citep{wiese+06}, so that was the initial interval chosen for all of them. 

For the remaining CE parameters, after many tests 
we adopted $k_{max}$~=~2000, $\alpha$ = 0.5, $q$ = 6, 
$N$=1000 and $N_{elite}$ = $0.05N$ for the calibration of each 1.5~\AA, which 
were the most efficient for the performance of the technique.

\section{Results}

ALiCCE begins the calibration by randomly choosing a value for $\log(gf)$, 
$\log~\Gamma_{StarK}$ and $\log \Gamma_{vdW}$ (within the limits of 400\%
of the original values) for each line in the given wavelength
interval, for each of the 1000 tentative solutions. This will create 1000 different
atomic line lists per iteration. For each of these lists, SYNTHE will generate a synthetic spectrum with
the atmospheric parameters of the Sun, Arcturus and Vega. These spectra will be 
compared with the simulated observed spectra and each list will be ranked by
their performance function (equation~\ref{perffunc}). New limits for the 
atomic parameters will be defined from the top 5\% of these lists
and the process will be repeated 2000 times. By the end of this process we 
expect to recover the atomic parameters used to generate the synthetic spectrum
with noise for the lines that are present with a reasonable strength in at least
one of the stars used in the calibration.

If a line is too weak or not even present in at least one of the stars used in the calibration, we
don't expect to recover its original parameters. Since we used synthetic spectra for 
this validation test we have prior information
about which lines are expected to be recovered by our code. When we produce synthetic spectra with SYNTHE, 
it generates an output in which it tabulates the contribution of each line in the spectra.
The numbers correspond to the per mil residual intensity at line center if the line were computed in isolation.
Table~\ref{participation} shows examples of these values for some of the lines in our interval. 
The closer the residual intensities are to 1.0, the weaker is the line. Lines weaker than $10^{-4}$ 
of the continuum will not be present in this output. Lines that appear as 1.0000 are rounded up values.

\begin{table}
\caption{Contribution of some lines in the spectra generated by SYNTHE.}
\begin {tabular} {@{}|c|ccccc}
\hline
$\lambda$ & Element  & Arcturus   & Sun  & Vega & Sum    \\
 (nm)     &         & Residual & Residual &  Residual   &  \\
\hline

851.3916 & ZrI &0.9936 & 0.9998 & 1.0000 & 2.9934 \\
851.4005 & FeII&0.9993 & 0.9978 & 0.9975 & 2.9946 \\
851.4072 & FeI &0.1365 & 0.3726 & 0.9931 & 1.5022 \\
851.4599 & SiI &0.8285 & 0.7805 & 0.9972 & 2.6062 \\
851.4629 & TiI &0.9998 & 1.0000 & 1.0000 & 2.9998 \\ 

\hline
\end{tabular}
\label{participation}
\end{table}

As a result of the present test, all lines where the sum in column 6 of table~\ref{participation}
is smaller than 2.800 had their $\log(gf)$ recovered 
(which means 8 lines in this
20~\AA~interval - central wavelengths, in nm: 
851.4072, 851.4599, 851.5109, 851.7085, 851.8028, 851.8352, 852.6628, 852.7198). For the broadening parameters
the lines have to be stronger. The $\log \Gamma_{vdW}$, important in cool stars, 
was only recovered for very strong lines in Arcturus
(Arcturus index in Table~\ref{participation} smaller than 0.2,
which means 2 lines in the 20~\AA~interval). The $\log \Gamma_{Stark}$ is more important in hot stars, so Vega was important here. Since in
the interval chosen there was no strong lines in Vega, $\log \Gamma_{Stark}$ 
did not converge for any line.

\begin{figure*} 
   \centering
   \includegraphics[width=17.5truecm]{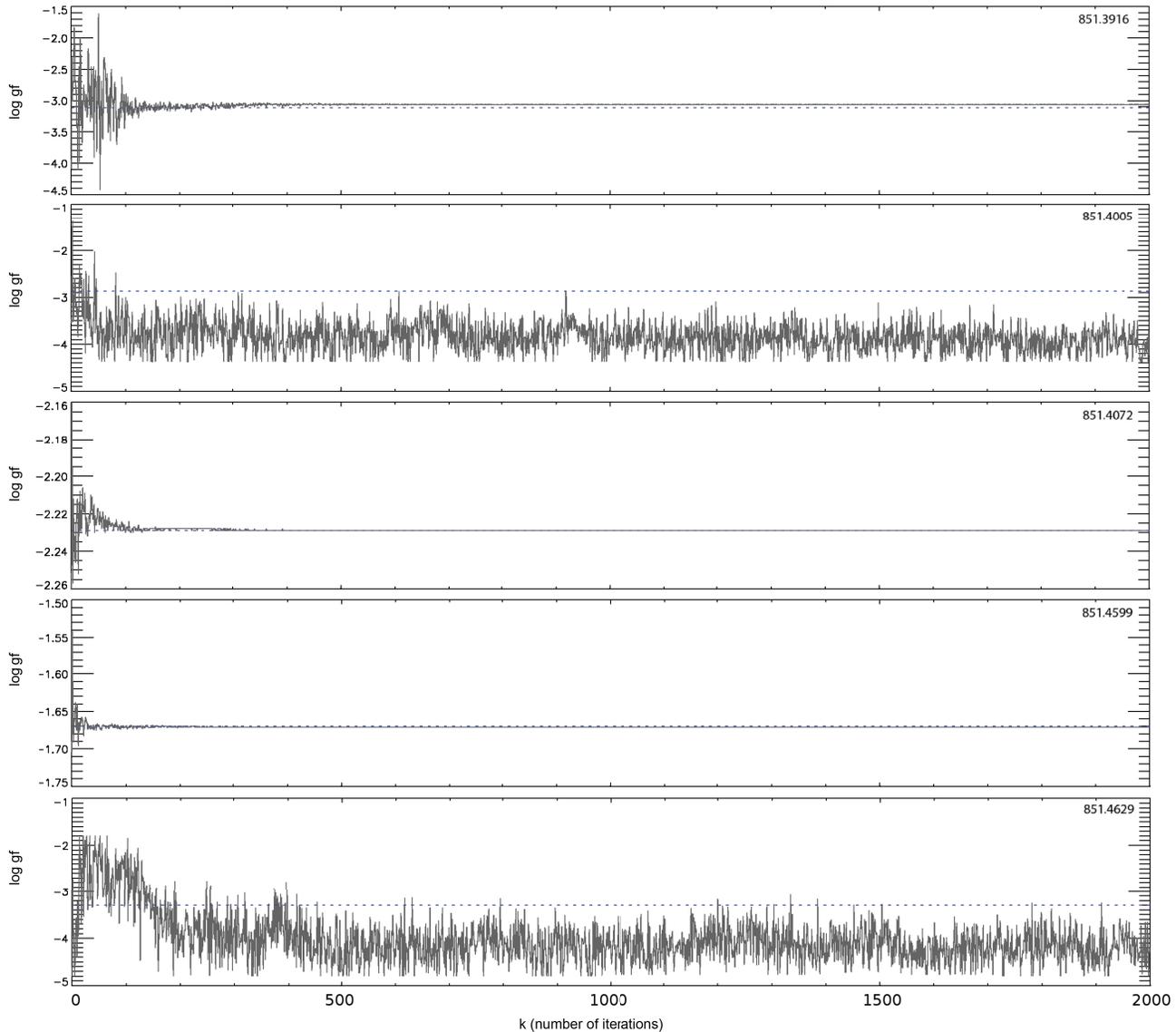} 
     \caption{Example of the evolution of the values of log gf found by ALiCCE at each iteration (black line) 
for the lines in table~\ref{participation}. The
doted blue line shows the expected value.}
\label{loggfconv}
\end{figure*}

As an example of the convergence behavior, figure~\ref{loggfconv} shows the  $\log(gf)$
recovery for the lines listed in Table~\ref{participation}. In this figure, the 
blue dotted line shows the expected value of $\log(gf)$, and the black line shows 
the evolution of the values found by ALiCCE as a function of iteration. The only
two lines that converged to the expected value in this figure are lines 851.4072 and 851.4599. 
However, line 851.3916 converged to a value which is not the nominal
value for this line. In this validation test we have this information, but when applying
the code for real stars, how will it be possible to distinguish between the 
correct values and the incorrect converged ones? It is clear that a simple visual inspection of 
figure~\ref{loggfconv} is not enough. 

\begin{figure} 
   \centering
   \includegraphics[width=8.6truecm]{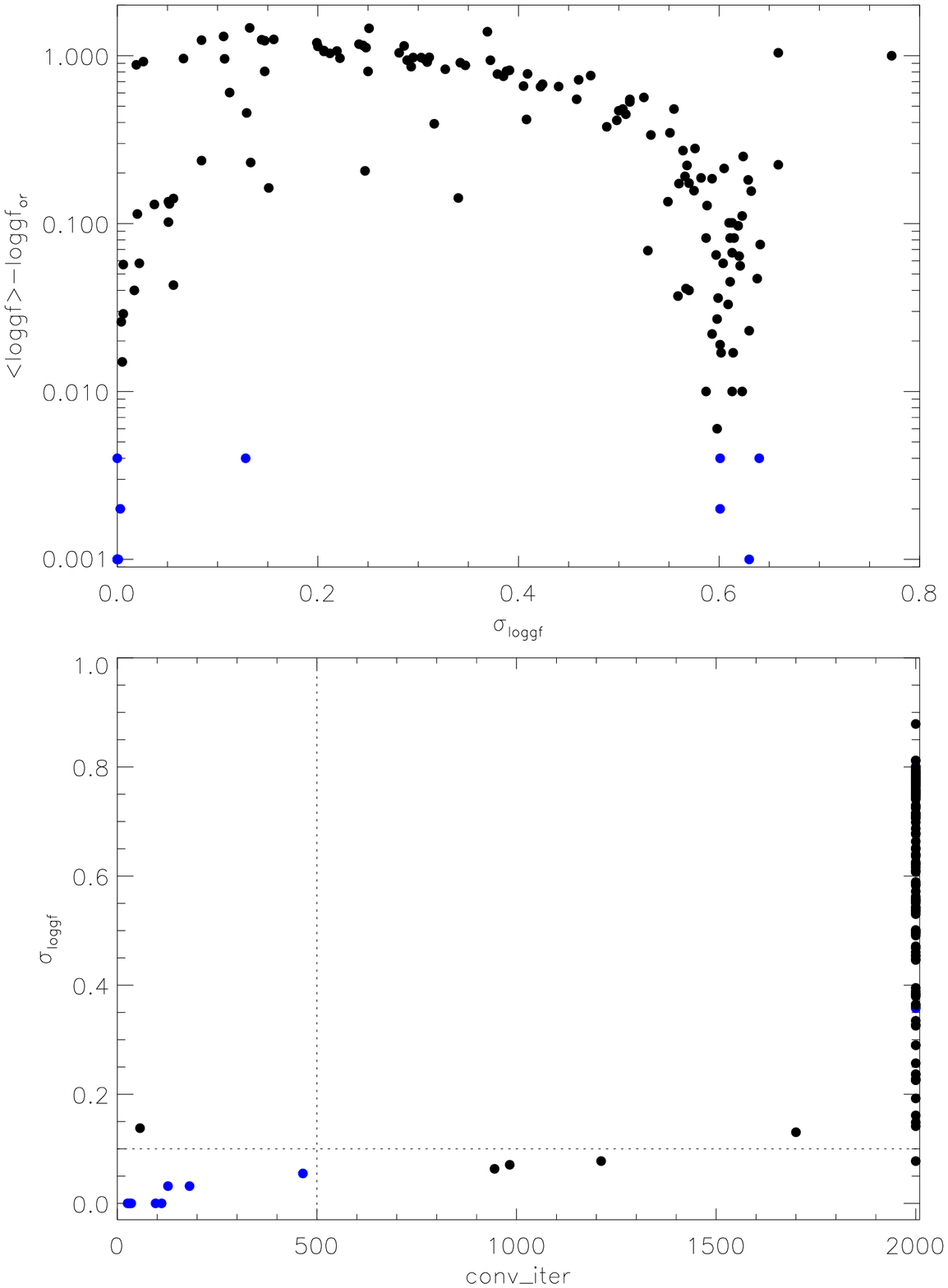} 
     \caption{Convergence criteria. The top panel shows the difference between expected values and
the values found by ALiCCE for each line in the interval, as a function of the errors calculated for
the calibration process. Points where $<\log(gf) >$ - $\log(gf_{or}) < 0.005$ 
were marked in blue.
The bottom panel shows the errors from the calibration process as a function
of the iteration in which the line converged to a single value. Blue points are the same as in the
top figure. The dotted
lines mark the criteria for convergence: values that converged in less than 500 iterations, 
and have uncertainties smaller than 0.1. Lines that
converged to the expected value are located in the bottom left of this figure.}
\label{convcrit}
\end{figure}

The criteria to select between the correct converged values and the incorrect ones are shown
in Figure~\ref{convcrit}. The top figure shows the difference between the average value 
of $\log(gf)$ found by ALiCCE ($<\log(gf) >$) and the original value used to generate the comparison
spectra ($\log(gf_{or})$) as a function of their uncertainties $\sigma_{\log(gf)}$,
as defined in Section 2.3.  
In this figure we show that
parameters that converged to their expected values have small uncertainties.
These are the lower left points (small $<\log(gf) >$ - $\log(gf_{or})$).
Some of these points are actually different points superimposed.
It is important to realize that not all points with small $<\log(gf) >$ - $\log(gf_{or})$ 
converged. Although their average value oscillates around the expected value,
they did not converge to this value, and thus, have larger uncertainties. We marked
all points values of $<\log(gf) >$ - $\log(gf_{or}) < 0.005$ in blue.
 
The bottom figure shows this effect more clearly. In this figure we show the uncertainties
in the $\log(gf)$ as a function of the iteration of the convergence, as defined in section 2.3. 
Blue points are the same as in the top
figure.
Lines that converged to their
expected values do so in the first iterations. Using this as a criteria, we can
identify lines that converged to an expected value even if we don't know
this value a priori, or have any knowledge of their contribution in the spectra of the stars. 
Based on our results we defined that lines that converged to the expected value do so
in less than 500 iterations, and have uncertainties smaller than 0.1. These limits
are marked by dotted lines in this figure. 
This is the main reason for the stopping criteria of the method to be the maximum number of 
iterations ($k_\rmn{max}$), and not a predefined value of convergence.
Given these results we believe that ALiCCE is an efficient tool for the calibration of atomic line lists 
for stellar spectrum synthesis. 

Amongst the parameters the code calibrates, $\log(gf)$ was the one with higher success of
recovery. This is not surprising as $\log(gf)$ usually is the main parameter (amongst the ones
studied) that drives the line profile. For the broadening parameters the recovery rate was
much lower. 
The recovery of the broadening parameters
is extremely dependent on the stellar rotational velocity. If the rotational velocity is
too high, these parameters cannot be recovered, since the Doppler broadening effect will
dominate the line profile. For the velocity of the stars
used in these calibrations we can recover the broadening parameter only for very strong
lines. 

\subsection{Results with observed spectra}

\begin{figure} 
   \centering
   \includegraphics[width=8.6truecm]{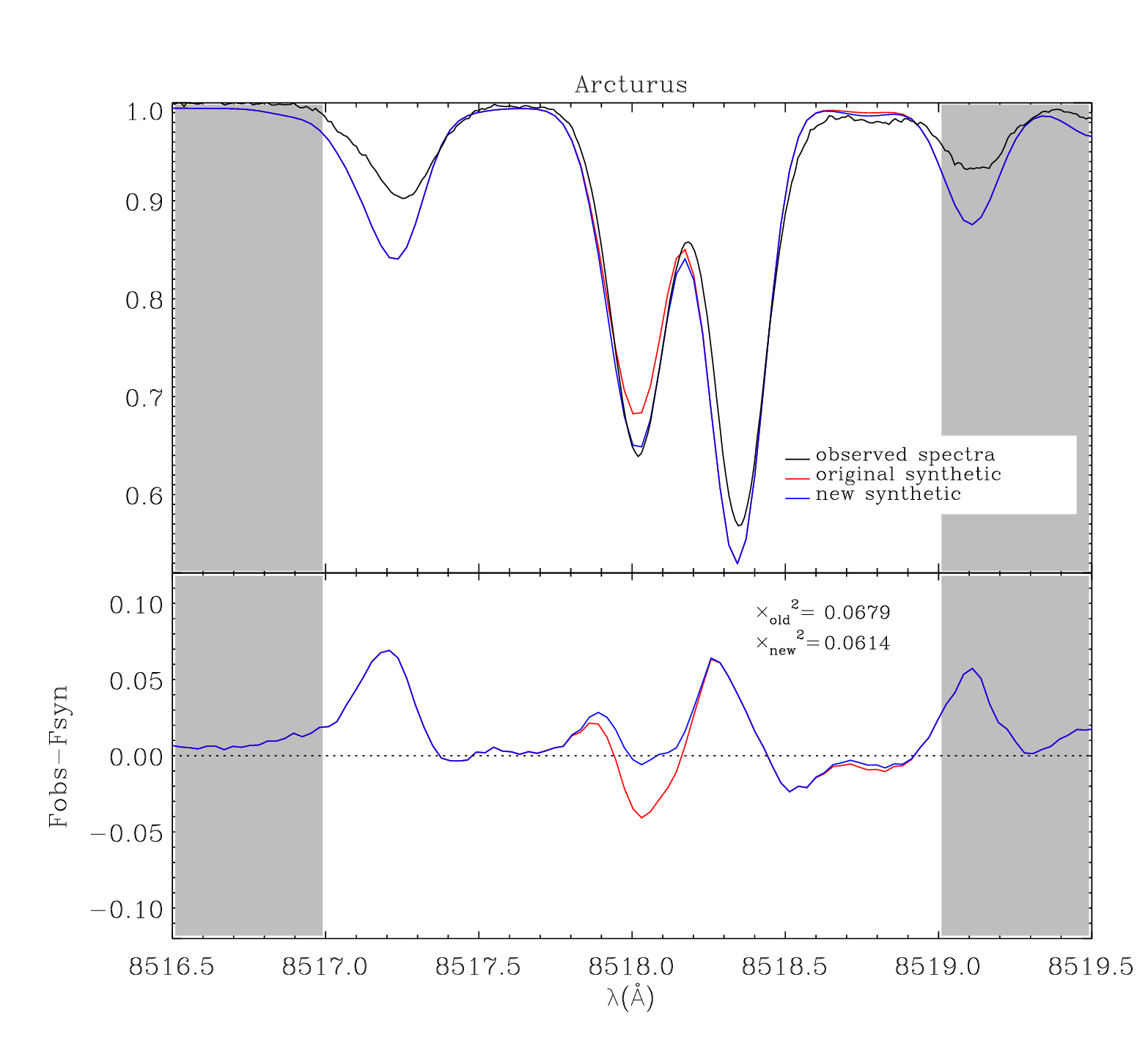} 
  \caption{Comparison between the synthetic spectra generated for Arcturus before (red line) and 
after (blue line) the calibration of the atomic line list by ALiCCE, using a real observed spectrum
(black line).}
\label{comp_obs_arc}
\end{figure}

\begin{figure} 
   \centering
   \includegraphics[width=8.6truecm]{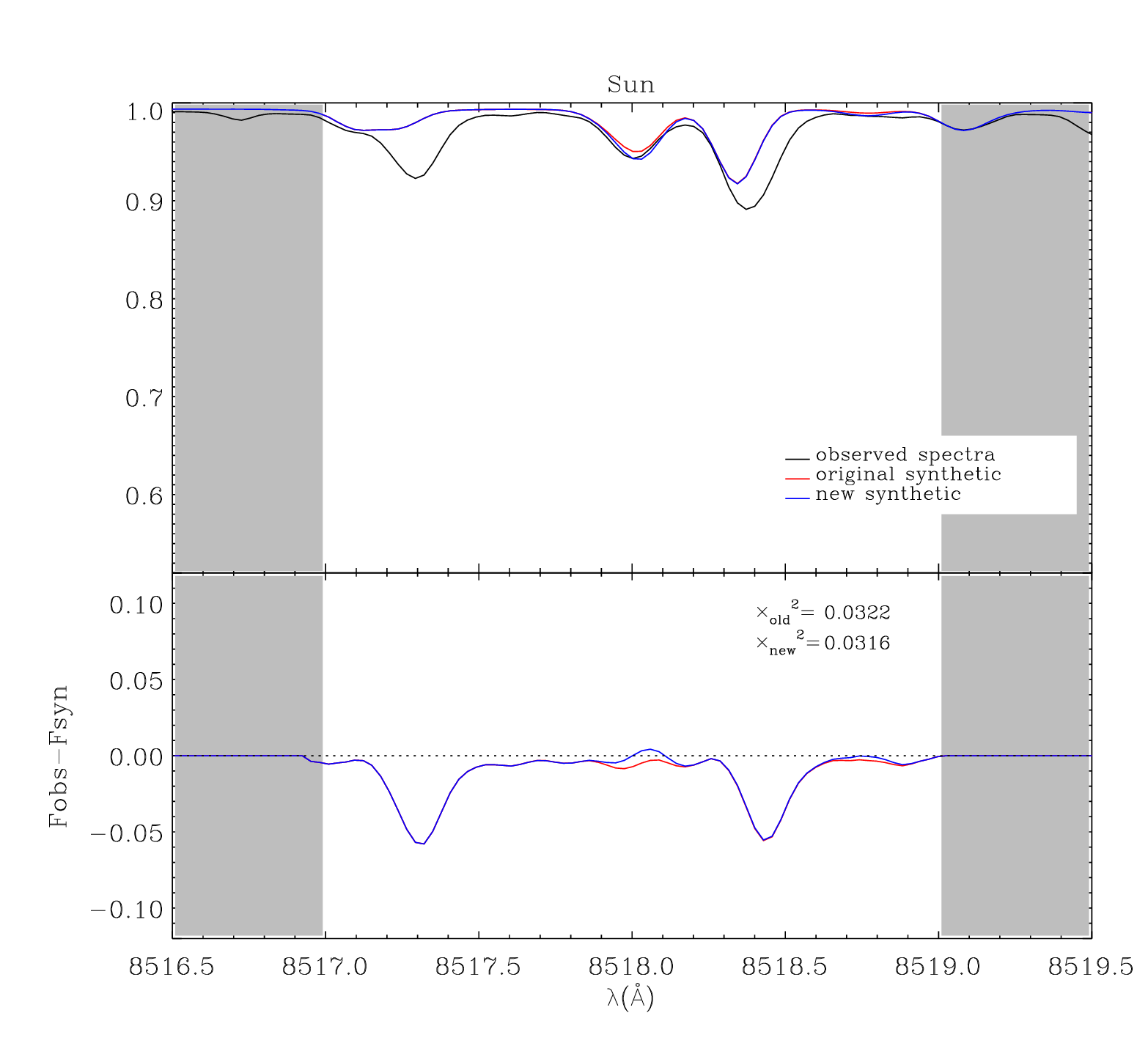} 
  \caption{Comparison between the synthetic spectra generated for the Sun before (red line) and 
after (blue line) the calibration of the atomic line list by ALiCCE, using a real observed spectrum
(black line).}
\label{comp_obs_sun}
\end{figure}

We performed a test with real data, applying the method to a small wavelength region from 8517 to 8519~\AA. Figures~\ref{comp_obs_arc} and~\ref{comp_obs_sun} show the results for the Sun and Arcturus respectively (Vega is almost featureless in this wavelength region, see Fig. 1).
The observed solar spectrum was obtained from \citet{kurucz04}, which is a revised solar flux atlas from \citet{kurucz+84}, and Arcturus spectrum from \citet{hinkle+00}.

This test illustrates the importance of calibrating the line list with more than one star: some
of the lines in the synthetic spectra are too weak for one star while too strong for the other. Calibrating
these lines for one of the stars alone would induce larger uncertainties in the other. 
In general, both spectra improved as shown by the values of the $\chi^2$ calculated from the difference between observed and synthetic spectra: 10\% improvement for Arcturus and 2\% for the Sun. The improvement of one line at 8518.028~\AA~ 
is particularly obvious. The non-fitted lines at 8517.30~\AA~ and  8518.45~\AA~ indicate possible missing lines in the atomic lines list.

A forthcoming paper will focus on the actual calibration of a larger wavelength window using new high-quality observed data, and with the inclusion of several missing lines.

\subsection{Physical Meaning}

Ideally, the calibration of atomic lines through this process could give us 
insights on
atomic physics and structure. 
This would be true if the physics and numerical approximations adopted in the
codes used to model atmospheres and spectra were valid,
and if the errors in the stellar atmospheric parameters (temperature, superficial gravity, chemical abundances, etc) were negligible.
 However, it is known that there are a 
lot of approximations adopted in the models (e.g. plane-parallel geometry, local thermodynamic
equilibrium, convection treatment, etc). This means that when an atomic line in a synthetic spectrum 
does not fit the observations, many other effects besides their imprecise atomic parameters might be
playing a role. 

Since we have no way of quantifying exactly how much each of these approximations affect the line
profiles, it is inevitable that when we calibrate the atomic line list using a given model atmosphere
and a given synthesis code the values found might sometimes compensate for these effects. What comes from this is that the interpretation
of the values found in terms of atomic physics might be compromised, and mainly, that the calibrated
line list created by ALiCCE should only be used for the same atmosphere models and synthesis codes for
which they were calibrated.

\section{Conclusion}

In this work we present the validation of the ALiCCE code for calibrating atomic line lists
used by spectral synthesis codes to generate stellar spectra
called ALiCCE, which uses the Cross-Entropy algorithm. For
validating our technique we simulated a calibration 
against synthetic spectra with noise of three well known stars with very well
determined atmospheric parameters: the Sun, Arcturus and Vega. 

The results obtained show that the Cross-Entropy algorithm is effective to
recover the atomic lines parameters used in line lists of spectral synthesis codes
for lines with  of the continuum flux.
The code was able to recover the $\log(gf)$ of the lines that had reasonable signal
in at least one of the three stars. In addition,
the code is able to recover broadening parameters of very strong lines, when 
they are present.

It is important to realize that the calibration of the atomic line list is tied
to a given model atmosphere and synthesis code. In this version of ALiCCE we used ATLAS9
model atmospheres and SYNTHE spectral code. Any line list calibrated with this version 
of ALiCCE should only be used for these codes.

\section*{Acknowledgments}

We thank Fiorella Castelli, the referee of this paper, for her valuable comments that
definetly improved the paper.
This research has been partially supported by the Brazilian agency FAPESP (2011/00171-4 and 2008/58406-4). 
L.M. also thanks L'Oreal Brasil and ABC for financial support, and CNPq through grant 305291/2012-2.
P.C. thanks CNPq throught grant 304291/2012-9.

\bsp

\label{lastpage}


\begin{thebibliography}{}

\bibitem[\protect\citeauthoryear{{Anstee} \& {O'Mara}}{{Anstee} \&
  {O'Mara}}{1995}]{anstee+95}
{Anstee} S.~D.,  {O'Mara} B.~J.,  1995, \mnras, 276, 859

\bibitem[\protect\citeauthoryear{{Bac{\l}awski}}{{Bac{\l}awski}}{2011}]{baclaw%
ski11}
{Bac{\l}awski} A.,  2011, European Physical Journal D, 61, 327

\bibitem[\protect\citeauthoryear{{Barbuy}, {Perrin}, {Katz}, {Coelho},
  {Cayrel}, {Spite} \& {Van't Veer-Menneret}}{{Barbuy}
  et~al.}{2003}]{barbuy+03}
{Barbuy} B.,  {Perrin} M.-N.,  {Katz} D.,  {Coelho} P.,  {Cayrel} R.,  {Spite}
  M.,    {Van't Veer-Menneret} C.,  2003, \aap, 404, 661

\bibitem[\protect\citeauthoryear{{Barklem}, {Anstee} \& {O'Mara}}{{Barklem}
  et~al.}{1998}]{barklem+98}
{Barklem} P.~S.,  {Anstee} S.~D.,    {O'Mara} B.~J.,  1998, \pasa, 15, 336

\bibitem[\protect\citeauthoryear{{Barklem} \& {O'Mara}}{{Barklem} \&
  {O'Mara}}{1997}]{barklem+97}
{Barklem} P.~S.,  {O'Mara} B.~J.,  1997, \mnras, 290, 102

\bibitem[\protect\citeauthoryear{{Barklem}, {Piskunov} \& {O'Mara}}{{Barklem}
  et~al.}{2000}]{barklem+00}
{Barklem} P.~S.,  {Piskunov} N.,    {O'Mara} B.~J.,  2000, \aaps, 142, 467

\bibitem[\protect\citeauthoryear{{Bell}, {Paltoglou} \& {Tripicco}}{{Bell}
  et~al.}{1994}]{bell+94}
{Bell} R.~A.,  {Paltoglou} G.,    {Tripicco} M.~J.,  1994, \mnras, 268, 771

\bibitem[\protect\citeauthoryear{{Blackwell-Whitehead}, {Pickering}, {Jones},
  {Nilsson} \& {Hartman}}{{Blackwell-Whitehead} et~al.}{2008}]{blackwell+08}
{Blackwell-Whitehead} R.~J.,  {Pickering} J.~C.,  {Jones} H.~R.~A.,  {Nilsson}
  H.,    {Hartman} H.,  2008, Journal of Physics Conference Series, 130, 012002

\bibitem[\protect\citeauthoryear{{Borrero}, {Bellot Rubio}, {Barklem} \& {del
  Toro Iniesta}}{{Borrero} et~al.}{2003}]{borrero+03}
{Borrero} J.~M.,  {Bellot Rubio} L.~R.,  {Barklem} P.~S.,    {del Toro Iniesta}
  J.~C.,  2003, \aap, 404, 749

\bibitem[\protect\citeauthoryear{{Caproni}, {Abraham} \& {Monteiro}}{{Caproni}
  et~al.}{2013}]{caproni+13}
{Caproni} A.,  {Abraham} Z.,    {Monteiro} H.,  2013, \mnras, 428, 280

\bibitem[\protect\citeauthoryear{{Caproni}, {Monteiro} \& {Abraham}}{{Caproni}
  et~al.}{2009}]{caproni+09}
{Caproni} A.,  {Monteiro} H.,    {Abraham} Z.,  2009, \mnras, 399, 1415

\bibitem[\protect\citeauthoryear{{Caproni}, {Monteiro}, {Abraham}, {Teixeira}
  \& {Toffoli}}{{Caproni} et~al.}{2011}]{caproni+11}
{Caproni} A.,  {Monteiro} H.,  {Abraham} Z.,  {Teixeira} D.~M.,    {Toffoli}
  R.~T.,  2011, \apj, 736, 68

\bibitem[\protect\citeauthoryear{{Castelli} \& {Kurucz}}{{Castelli} \&
  {Kurucz}}{1994}]{Castelli+94}
{Castelli} F.,  {Kurucz} R.~L.,  1994, \aap, 281, 817

\bibitem[\protect\citeauthoryear{{Castelli} \& {Kurucz}}{{Castelli} \&
  {Kurucz}}{2004}]{castelli+04}
{Castelli} F.,  {Kurucz} R.~L.,  2004, \aap, 419, 725

\bibitem[\protect\citeauthoryear{{Civi{\v s}}, {Ferus}, {Kubel{\'{\i}}k},
  {Jelinek}, {Chernov} \& {Zanozina}}{{Civi{\v s}} et~al.}{2012}]{civis+12}
{Civi{\v s}} S.,  {Ferus} M.,  {Kubel{\'{\i}}k} P.,  {Jelinek} P.,  {Chernov}
  V.~E.,    {Zanozina} E.~M.,  2012, \aap, 542, A35

\bibitem[\protect\citeauthoryear{{de Boer}, {Kroese}, {Mannor} \&
  {Rubinstein}}{{de Boer} et~al.}{2005}]{deboer+05}
{de Boer} P.~T.,  {Kroese} D.~P.,  {Mannor} S.,    {Rubinstein} R.~Y.,  2005,
  Annals of Operatons Research, 134, 19

\bibitem[\protect\citeauthoryear{{Den Hartog}, {Lawler}, {Sobeck}, {Sneden} \&
  {Cowan}}{{Den Hartog} et~al.}{2011}]{denHartog+11}
{Den Hartog} E.~A.,  {Lawler} J.~E.,  {Sobeck} J.~S.,  {Sneden} C.,    {Cowan}
  J.~J.,  2011, \apjs, 194, 35

\bibitem[\protect\citeauthoryear{{Derouich}, {Sahal-Br{\'e}chot}, {Barklem} \&
  {O'Mara}}{{Derouich} et~al.}{2003}]{derouich+03}
{Derouich} M.,  {Sahal-Br{\'e}chot} S.,  {Barklem} P.~S.,    {O'Mara} B.~J.,
  2003, \aap, 404, 763

\bibitem[\protect\citeauthoryear{{Dimitrijevi{\'c}}, {Ryabchikova},
  {Popovi{\'c}}, {Shulyak} \& {Tsymbal}}{{Dimitrijevi{\'c}}
  et~al.}{2003}]{dimitrijevic+03}
{Dimitrijevi{\'c}} M.~S.,  {Ryabchikova} T.,  {Popovi{\'c}} L.~{\v C}.,
  {Shulyak} D.,    {Tsymbal} V.,  2003, \aap, 404, 1099

\bibitem[\protect\citeauthoryear{{Fuhr} \& {Wiese}}{{Fuhr} \&
  {Wiese}}{2006}]{fuhr+06}
{Fuhr} J.~R.,  {Wiese} W.~L.,  2006, J.~Phys.~Chem.~Ref.~Data., 35, 1669

\bibitem[\protect\citeauthoryear{{Gray}}{{Gray}}{1981}]{gray81}
{Gray} D.~F.,  1981, \apj, 251, 155

\bibitem[\protect\citeauthoryear{{Hinkle}, {Wallace}, {Valenti} \&
  {Harmer}}{{Hinkle} et~al.}{2000}]{hinkle+00}
{Hinkle} K.,  {Wallace} L.,  {Valenti} J.,    {Harmer} D.,  2000, {Visible and
  Near Infrared Atlas of the Arcturus Spectrum 3727-9300 A}

\bibitem[\protect\citeauthoryear{{Jorissen}}{{Jorissen}}{2004}]{jorissen04}
{Jorissen} A.,  2004, Physica Scripta Volume T, 112, 73

\bibitem[\protect\citeauthoryear{{Klose}, {Fuhr} \& {Wiese}}{{Klose}
  et~al.}{2002}]{klose+02}
{Klose} J.~Z.,  {Fuhr} J.~R.,    {Wiese} W.~L.,  2002, Journal of Physical and
  Chemical Reference Data, 31, 217

\bibitem[\protect\citeauthoryear{{Konjevi{\'c}}, {Lesage}, {Fuhr} \&
  {Wiese}}{{Konjevi{\'c}} et~al.}{2002}]{konjevic+02}
{Konjevi{\'c}} N.,  {Lesage} A.,  {Fuhr} J.~R.,    {Wiese} W.~L.,  2002,
  Journal of Physical and Chemical Reference Data, 31, 819

\bibitem[\protect\citeauthoryear{{Kroese}, {Porotsky} \& {Rubinstein}}{{Kroese}
  et~al.}{2006}]{kroese+06}
{Kroese} D.~P.,  {Porotsky} S.,    {Rubinstein} R.~Y.,  2006, Methodol. Comput.
  Appl. Probab., 8, 383

\bibitem[\protect\citeauthoryear{{Kurucz}}{{Kurucz}}{1970}]{kurucz70}
{Kurucz} R.~L.,  1970, SAO Special Report, 309

\bibitem[\protect\citeauthoryear{{Kurucz}}{{Kurucz}}{1992}]{kurucz92}
{Kurucz} R.~L.,  1992, \rmxaa, 23, 45

\bibitem[\protect\citeauthoryear{{Kurucz}}{{Kurucz}}{2005}]{kurucz04}
{Kurucz} R.~L.,  2005, Memorie della Societa Astronomica Italiana Supplementi,
  8, 189

\bibitem[\protect\citeauthoryear{{Kurucz}}{{Kurucz}}{2011}]{kurucz11}
{Kurucz} R.~L.,  2011, Canadian Journal of Physics, 89, 417

\bibitem[\protect\citeauthoryear{{Kurucz} \& {Avrett}}{{Kurucz} \&
  {Avrett}}{1981}]{kurucz+81}
{Kurucz} R.~L.,  {Avrett} E.~H.,  1981, SAO Special Report, 391

\bibitem[\protect\citeauthoryear{{Kurucz}, {Furenlid}, {Brault} \&
  {Testerman}}{{Kurucz} et~al.}{1984}]{kurucz+84}
{Kurucz} R.~L.,  {Furenlid} I.,  {Brault} J.,    {Testerman} L.,  1984, {Solar
  flux atlas from 296 to 1300 nm}

\bibitem[\protect\citeauthoryear{{Lesage}, {Konjevic} \& {Fuhr}}{{Lesage}
  et~al.}{1999}]{lesage+99}
{Lesage} A.,  {Konjevic} N.,    {Fuhr} J.~R.,  1999, in {Herman} R.~M.,  ed.,
  Spectral Line Shapes Vol.~467 of American Institute of Physics Conference
  Series, {Progress in spectral line shapes and shifts evaluation of
  experimental Stark broadening parameters}.
pp 27--36

\bibitem[\protect\citeauthoryear{{Margolin}}{{Margolin}}{2004}]{margolin04}
{Margolin} L.,  2004, Annals of Operations Research, 134, 201

\bibitem[\protect\citeauthoryear{{Martins} \& {Coelho}}{{Martins} \&
  {Coelho}}{2007}]{martins+07}
{Martins} L.~P.,  {Coelho} P.,  2007, \mnras, 381, 1329

\bibitem[\protect\citeauthoryear{{Mel{\'e}ndez} \& {Barbuy}}{{Mel{\'e}ndez} \&
  {Barbuy}}{2009}]{melendez+09}
{Mel{\'e}ndez} J.,  {Barbuy} B.,  2009, \aap, 497, 611

\bibitem[\protect\citeauthoryear{{Mel{\'e}ndez}, {Barbuy}, {Bica}, {Zoccali},
  {Ortolani}, {Renzini} \& {Hill}}{{Mel{\'e}ndez} et~al.}{2003}]{melendez+03}
{Mel{\'e}ndez} J.,  {Barbuy} B.,  {Bica} E.,  {Zoccali} M.,  {Ortolani} S.,
  {Renzini} A.,    {Hill} V.,  2003, \aap, 411, 417

\bibitem[\protect\citeauthoryear{{Monteiro}, {Dias} \& {Caetano}}{{Monteiro}
  et~al.}{2010}]{monteiro+10}
{Monteiro} H.,  {Dias} W.~S.,    {Caetano} T.~C.,  2010, \aap, 516, A2

\bibitem[\protect\citeauthoryear{{Munari}, {Sordo}, {Castelli} \&
  {Zwitter}}{{Munari} et~al.}{2005}]{munari+05}
{Munari} U.,  {Sordo} R.,  {Castelli} F.,    {Zwitter} T.,  2005, \aap, 442,
  1127

\bibitem[\protect\citeauthoryear{{Oliveira}, {Monteiro}, {Dias} \&
  {Caetano}}{{Oliveira} et~al.}{2013}]{oliveira+13}
{Oliveira} A.~F.,  {Monteiro} H.,  {Dias} W.~S.,    {Caetano} T.~C.,  2013,
  \aap, 557, A14

\bibitem[\protect\citeauthoryear{{Peterson}, {Hummel}, {Pauls}, {Armstrong},
  {Benson}, {Gilbreath}, {Hindsley}, {Hutter}, {Johnston}, {Mozurkewich} \&
  {Schmitt}}{{Peterson} et~al.}{2006}]{peterson+06}
{Peterson} D.~M.,  {Hummel} C.~A.,  {Pauls} T.~A.,  {Armstrong} J.~T.,
  {Benson} J.~A.,  {Gilbreath} G.~C.,  {Hindsley} R.~B.,  {Hutter} D.~J.,
  {Johnston} K.~J.,  {Mozurkewich} D.,    {Schmitt} H.~R.,  2006, \nat, 440,
  896

\bibitem[\protect\citeauthoryear{{Pickering}, {Blackwell-Whitehead}, {Thorne},
  {Ruffoni} \& {Holmes}}{{Pickering} et~al.}{2011}]{pickering+11}
{Pickering} J.~C.,  {Blackwell-Whitehead} R.,  {Thorne} A.~P.,  {Ruffoni} M.,
   {Holmes} C.~E.,  2011, Canadian Journal of Physics, 89, 387

\bibitem[\protect\citeauthoryear{{Rubinstein}}{{Rubinstein}}{1997}]{rubinstein%
97}
{Rubinstein} R.~Y.,  1997, European Journal of Operational Research, 99, 89

\bibitem[\protect\citeauthoryear{{Rubinstein}}{{Rubinstein}}{1999}]{rubinstein%
99}
{Rubinstein} R.~Y.,  1999, Methodology and Computing in Applied Probability, 2,
  127

\bibitem[\protect\citeauthoryear{{Ruffoni}, {Allende Prieto}, {Nave} \&
  {Pickering}}{{Ruffoni} et~al.}{2013}]{ruffoni+13}
{Ruffoni} M.~P.,  {Allende Prieto} C.,  {Nave} G.,    {Pickering} J.~C.,  2013,
  \apj, 779, 17

\bibitem[\protect\citeauthoryear{{Safronova} \& {Safronova}}{{Safronova} \&
  {Safronova}}{2010}]{safronova+10}
{Safronova} U.~I.,  {Safronova} M.~S.,  2010, Journal of Physics B Atomic
  Molecular Physics, 43, 074025

\bibitem[\protect\citeauthoryear{{Sbordone}, {Bonifacio}, {Castelli} \&
  {Kurucz}}{{Sbordone} et~al.}{2004}]{sbordone+04}
{Sbordone} L.,  {Bonifacio} P.,  {Castelli} F.,    {Kurucz} R.~L.,  2004,
  Memorie della Societa Astronomica Italiana Supplementi, 5, 93

\bibitem[\protect\citeauthoryear{{Shchukina} \& {Vasil'eva}}{{Shchukina} \&
  {Vasil'eva}}{2013}]{shchukina+13}
{Shchukina} N.~G.,  {Vasil'eva} I.~E.,  2013, Kinematics and Physics of
  Celestial Bodies, 29, 53

\bibitem[\protect\citeauthoryear{{Short} \& {Lester}}{{Short} \&
  {Lester}}{1996}]{short+96}
{Short} C.~I.,  {Lester} J.~B.,  1996, \apj, 469, 898

\bibitem[\protect\citeauthoryear{{Smith}}{{Smith}}{1978}]{smith78}
{Smith} M.~A.,  1978, \apj, 224, 584

\bibitem[\protect\citeauthoryear{{Sobeck}, {Lawler} \& {Sneden}}{{Sobeck}
  et~al.}{2007}]{sobeck+07}
{Sobeck} J.~S.,  {Lawler} J.~E.,    {Sneden} C.,  2007, \apj, 667, 1267

\bibitem[\protect\citeauthoryear{{Stalin}, {Trivedi}, {Sinha} \&
  {Sanwal}}{{Stalin} et~al.}{1997}]{stalin+97}
{Stalin} C.~S.,  {Trivedi} C.,  {Sinha} K.,    {Sanwal} B.~B.,  1997, Bulletin
  of the Astronomical Society of India, 25, 353

\bibitem[\protect\citeauthoryear{{Taklif}}{{Taklif}}{1990}]{taklif90}
{Taklif} A.~G.,  1990, \physscr, 42, 69

\bibitem[\protect\citeauthoryear{{Wahlgren}}{{Wahlgren}}{2010}]{wahlgren10}
{Wahlgren} G.~M.,  2010, in {Monier} R.,  {Smalley} B.,  {Wahlgren} G.,
  {Stee} P.,  eds, EAS Publications Series Vol.~43 of EAS Publications Series,
  {Oscillator Stengths and Their Uncertainties}.
pp 91--114

\bibitem[\protect\citeauthoryear{{Wiese} \& {Fuhr}}{{Wiese} \&
  {Fuhr}}{2006}]{wiese+06}
{Wiese} W.~L.,  {Fuhr} J.~R.,  2006, in {Weck} P.~F.,  {Kwong} V.~H.~S.,
  {Salama} F.,  eds, NASA LAW 2006 {New Critical Compilations of Atomic
  Transition Probabilities for Neutral and Singly Ionized Carbon, Nitrogen, and
  Iron}.
p.~278

\bibitem[\protect\citeauthoryear{{Wiese}, {Fuhr} \& {Bridges}}{{Wiese}
  et~al.}{2011}]{wiese+11}
{Wiese} W.~L.,  {Fuhr} J.~R.,    {Bridges} J.~M.,  2011, in 2010 NASA
  Laboratory Astrophysics Workshop {Towards More Accurate Atomic Oscillator
  Strengths}.
p.~C16

\bibitem[\protect\citeauthoryear{{Wood}, {Lawler}, {Sneden} \& {Cowan}}{{Wood}
  et~al.}{2013}]{wood+13}
{Wood} M.~P.,  {Lawler} J.~E.,  {Sneden} C.,    {Cowan} J.~J.,  2013, \apjs,
  208, 27


\end{thebibliography}
\end{document}